\documentclass[aps,pra,twocolumn,superscriptaddress,showpacs]{revtex4-1}
\usepackage[utf8]{inputenc} 

\usepackage{amssymb,amsmath}
\usepackage{graphics}
\usepackage{xcolor}
\usepackage{dsfont}
\usepackage{comment}

\usepackage{textgreek}
\usepackage{physics}
\usepackage{graphicx}

\def\ex{\mathbf{e}_x}  
\def\ey{\mathbf{e}_y}  
\def\ez{\mathbf{e}_z}  

\def\kr{k_{\rm R}}                            				
\def\vr{v_{\rm R}}                            				
\def\Er{E_{\rm R}} 
\def\lambdar{\lambda_{\rm R}} 
\def\Or{\Omega_{\rm R}}

\usepackage{etoolbox}
\newtoggle{SUPPORT}
\toggletrue{SUPPORT}
\togglefalse{SUPPORT}
\long\def\support#1{ \iftoggle{SUPPORT}{\begin{widetext}\small {\leavevmode\color{red}{#1}} \normalsize\end{widetext}}{} }

\begin{document}

\title{Enhanced transport of spin-orbit coupled Bose gases in disordered potentials}

\author{Y. Yue}
\affiliation{Joint Quantum Institute, National Institute of Standards and Technology, and University of Maryland, Gaithersburg, Maryland, 20899, USA}
\author{C. A. R. Sá de Melo}
\affiliation{School of Physics, Georgia Institute of Technology, Atlanta, Georgia 30332, USA}
\author{I. B. Spielman}
\affiliation{Joint Quantum Institute, National Institute of Standards and Technology, and University of Maryland, Gaithersburg, Maryland, 20899, USA}
\email{ian.spielman@nist.gov}
\homepage{http://ultracold.jqi.umd.edu}
\date{\today}

\begin{abstract}
Anderson localization is a single particle localization phenomena in disordered media that is accompanied by an absence of diffusion.
Spin-orbit coupling (SOC) describes an interaction between a particle's spin and its momentum that directly affects its energy dispersion, for example creating dispersion relations with gaps and multiple local minima.
We show theoretically that combining one-dimensional spin-orbit coupling with a transverse Zeeman field suppresses the effects of disorder, thereby increasing the localization length and conductivity.
This increase results from a suppression of back scattering between states in the gap of the SOC dispersion relation.
Here, we focus specifically on the interplay of disorder from an optical speckle potential and SOC generated by two-photon Raman processes in quasi-1D Bose-Einstein condensates.
We first describe back-scattering using a Fermi's golden rule approach, and then numerically confirm this picture by solving the time-dependent 1D Gross Pitaevskii equation for a weakly interacting Bose-Einstein condensate with SOC and disorder.
We find that on the 10's of millisecond time scale of typical cold atom experiments moving in harmonic traps, initial states with momentum in the zero-momentum SOC gap evolve with negligible back-scattering, while without SOC these same states rapidly localize.
\end{abstract}

\maketitle
Anderson Localization (AL), introduced in 1958~\cite{AL}, describes the localization of waves in disordered media. Anderson studied the evolution of a wave packet undergoing multiple scattering processes from a random potential and proved the scattered waves can constructively interfere, leading to localization. This general starting point makes AL applicable to many systems including: optical waves in disordered media~\cite{wiersma1997localization,scheffold1999localization,storzer2006observation}, electrons in imperfect crystals~\cite{AL} and matter waves in disordered optical potentials~\cite{sanchez2007anderson,billy2008direct,roati2008anderson}. In materials, microscopic electron scattering processes partly govern the macroscopic conductivity and AL predicts a metal-insulator transition. Increasing a system's conductivity therefore requires some change in these scattering processes.  The most straightforward mechanism is to reduce the disorder strength. Here we describe an alternate approach in which spin-orbit coupling (SOC) greatly suppresses the back scattering and thereby increases the conductivity.  We then propose a realization of this effect using a cold-atom Bose-Einstein condensate (BEC) with laser-induced SOC~\cite{lin2011spin} and disorder from optical speckle.

SOC is a ubiquitous phenomenon in physical systems that describes the interaction between a particle's spin and its momentum. When SOC is combined with a transverse magnetic field (in the sense of Zeeman shifts, not Lorentz forces), gaps in the dispersion relation can open at spin-degeneracy points.
The opening of these gaps modifies the electrons' scattering processes and affects transport.
AL was first realized for ultracold atomic systems~\cite{billy2008direct,roati2008anderson} in 2008, and the experimental techniques are now well established. 
Shortly thereafter,  techniques for creating SOC in the cold atom lab were demonstrated~\cite{lin2011spin}.
Together, this makes cold-atom systems an ideal platform to study the interplay between AL and SOC. 

Optical speckle is a powerful tool for creating disordered potentials for atomic systems~\cite{clement2006experimental}.  The strength of the resultant potential is under direct experimental control: the spatial correlation length is tunable and the correlation function is well known.  Here we analytically and numerically study backscattering in speckle potentials of quasi-1d spin-orbit coupled BECs (SOBECs) and compare to the case without SOC. We show that SOC can reduce the scattering processes for specific momentum states. In the broader context, our results suggest that in thin nano-wires, SOC might significantly decrease resistance and improve energy efficiency in electronic devices.

This manuscript is organized as follows. In Sec.~\ref{characteristics}, we begin with an introduction to optical speckle as it pertains to our proposal. In Sec.~II, we analytically calculate the probability an initial momentum state being scattered by the speckle potential to any final momentum state and show that SOC can reduce back scattering. Lastly in Sec.~III, we describe numerical simulations of quasi-1d BECs starting in different momentum states subject to a speckle potential with and without SOC. We show that even with the higher order scattering processes and interaction between particles present in the numerical simulations, SOC can reduce the localization effects of disorder and enhance transport. 

\section{Characterics of optical speckle}\label{characteristics}

Optical speckle can be understood as the self-interfering wave field of a laser after acquiring random phase by reflection off rough surfaces or transmission through disordered media, called a diffuser~\cite{goodman2007speckle}. We will focus on the transmission case and assume that the spatial scale of the disorder $\sigma$ is small in comparison to the laser beam size and that the diffuser transmits light uniformly. The transmitted field can be intuitively thought of as of many waves scattered from microscopic elements comprising the diffuser.  So randomness arises.  As a disordered field, optical speckle is characterized by its intensity distribution, spatial intensity correlation function and power spectral density (PSD).  

As shown in Fig.~\ref{fig:speckle_model} shows, ray optics in the paraxial limit provides a simple and useful approach to estimating the on-axis beam properties of a speckle beam a distance $z$ beyond a diffuser.  As a collimated laser beam of wavelength $\lambda$ travels through a diffuser of diameter $D_d$, it acquires a local divergence angle $\theta_d\simeq \lambda / (2 \sigma)$.   

Figure~\ref{fig:speckle_model}a depicts the most simple case consisting of an isolated diffuser, for which there are two qualitatively different regimes: A near-field regime with $z < D_d / (2 \theta_d)$, where the typical length scale of optical speckle is $\sigma$, and a far-field regime where the NA of the diffuser increases the speckle scale to $(\lambda / 2)\times(D_d / 2 z)$.  This simple approach is insufficient because we will be interested in micrometer scale speckle, which is far smaller than the 10 to 100 micrometer scale of $\sigma$ for commercial  diffusers.

In Fig.~\ref{fig:speckle_model}b we add a lens with diameter $D_L$ and focal length $f$  just after the diffuser.  In the focal plane of the lens, the speckle scale is set by the lens NA, giving a speckle length scale $\lambda f / D_L$, independent of $\sigma$. In contrast, the beam width at the focal plane $w(f) \simeq 2 f \theta_d$ is set by the speckle scale $\sigma$ and not the lens diameter.

In the following section we will derive the origin of these design guidelines from  the paraxial wave equation. 

\subsection{Gaussian beam equations with speckle}

\begin{figure}[htbp]
    \begin{center}
        \includegraphics{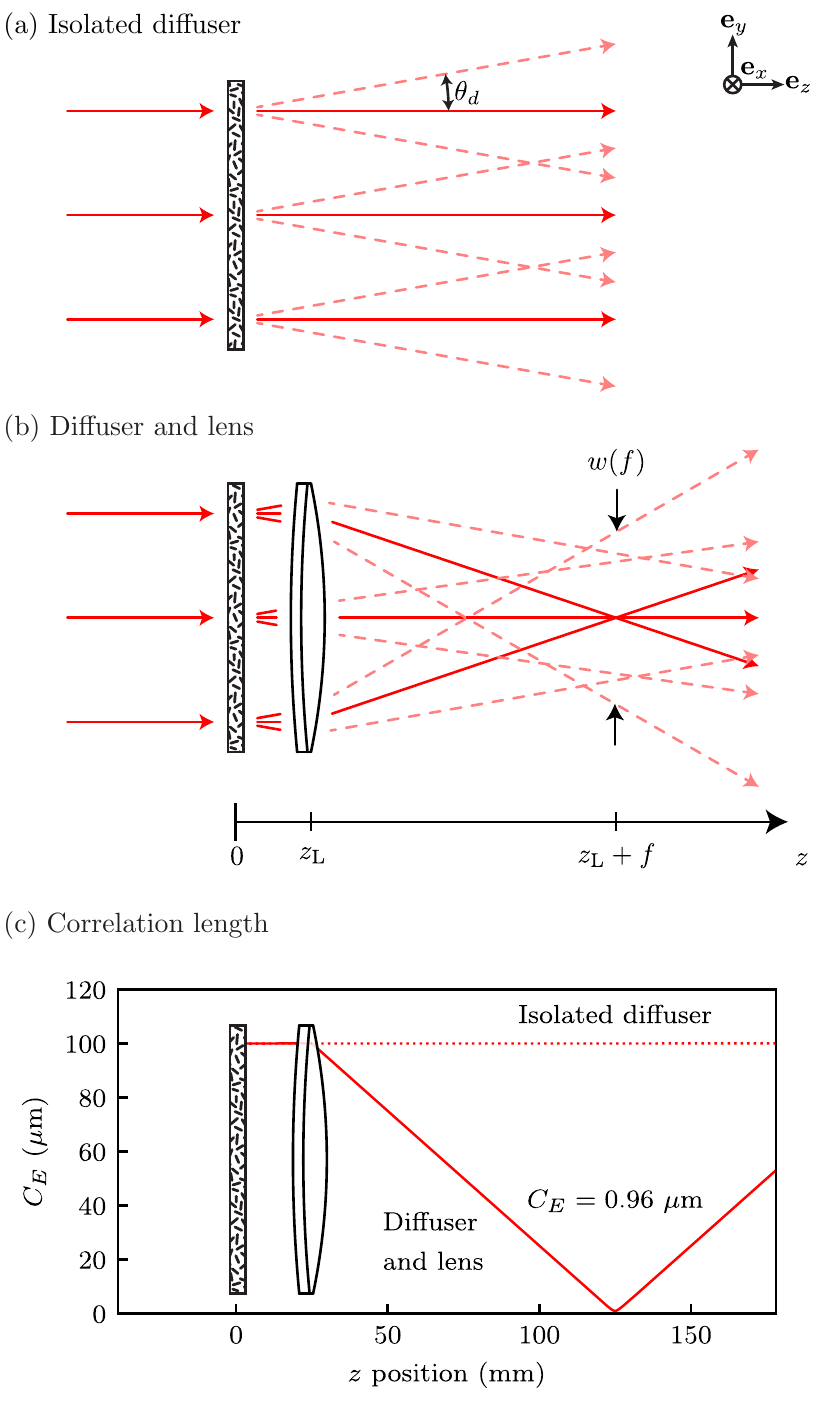}
    \end{center}

    \caption{Optical speckle schematic. (a) A collimated beam is transmitted through a rough medium and its intensity is measured in plane $z$. (b) The diverged beam after the rough medium is imaged by a lens at plane $z=z_L$ and $f$ is the focal point of the lens. (c) Field-field correlation length for a Gaussian speckle beam initially with $\sigma=100\ \mu{\rm m}$ and $w=25\ {\rm mm}$ as a function of propagation distance. The red curves plot $c_E(z)$ computed with (solid) and without (dashed) a lens with focal length $f=100\ {\rm mm}$ at $z_L=25\ {\rm mm}$. 
    }
    \label{fig:speckle_model}
\end{figure}

We focus on monochromatic optical electric fields $E({\bf x}, t)$ with angular frequency $\omega$ traveling predominantly along ${\bf e}_z$.  Such waves can be decomposed as $E({\bf x}, t) = E_\perp({\bf r}; z) \exp[i(k_0 z -\omega t)]$, where $E_\perp({\bf r}; z)$ describes the transverse structure of the electric field with the high spatial frequencies associated with the nominal propagation along ${\bf e}_z$ factored out.  For spatial scales in excess of the optical wavelength the transverse field obeys the paraxial wave equation
\begin{equation}\label{para equation}
    -2i k_0 \partial_z E_\perp({\bf r}; z) = \left[-\nabla_\perp^2 + k_0^2 \chi({\bf r}; z)\right] E_\perp({\bf r}; z)
\end{equation}
traveling in a material with relative susceptibility $\chi({\bf r}; z)$.  We will suppress the $\perp$ subscript in the remainder of our discussion.

Upon traversing through a thin but disordered material with susceptibility $\chi({\bf r}) $ and thickness $\delta z$, an initially Gaussian wave field $E^-({\bf r}, 0) = E_0\exp{-{\bf r}^2/w^2}$ acquires a position dependent complex phase $\phi({\bf r}) = \chi({\bf r}) k_0 \delta z/2$.  The resultant field
\begin{equation}
    E^+({\bf r}, 0) = E^-({\bf r}, 0)\exp[- i \phi({\bf r})]
\end{equation}
carries the imprint of the disordered medium. The field a distance $z$ beyond the speckle plate follows from
\begin{align}\label{para field}
    E({\bf r}; z) &= \frac{-i k_0}{2\pi z}\int d^2{\bf r'} E^+({\bf r'}; 0) e^{-i k_0|{\bf r}-{\bf r'}|^2/2z},
\end{align} 
the formal solution to the paraxial wave equation Eq.~(\ref{para equation}).  We model typical diffusion plates, for which: (1) the correlation function of the susceptibility $\langle \chi({\bf r_1})\chi({\bf r_2}) \rangle$ depends only on relative distance $|{\bf r_1}-{\bf r_2}|$, where $\langle ... \rangle$ denotes the ensemble average over disorder realizations. (2) the variation of the imprinted phase $\phi({\bf r})$ is much larger than $2\pi$ with
\begin{align}\label{eq:zeromean}
\langle \exp\left[- i \phi({\bf r}_1)\right]\rangle &= 0,
\end{align} 
i.e., $\phi({\bf r})$ is uniformly distributed over the interval $[-\pi,\pi]$.
 
We turn to the field-field correlation function
\begin{equation}\label{C_E}
\begin{aligned}
    C_E({\bf r}_1,{\bf r}_2;z) &=\langle E({\bf r}_1; z)E^\ast({\bf r}_2; z)\rangle\! -\! \langle E({\bf r}_1; z)\rangle\langle E^\ast({\bf r}_2; z)\rangle
\end{aligned}
\end{equation}
to characterize the statistical properties of the disordered electric field.  Equation~\eqref{eq:zeromean} implies that the second term is zero.  At $z=0$, the uniform phase distribution implies $\langle E^+({\bf r}; 0) \rangle = 0$, giving
\begin{align*}
\frac{C_E({\bf r}_1,{\bf r}_2;0)}{E_0^2 } &= \exp\left(-\frac{{\bf r}_1^2 + {\bf r}_2^2}{w^2}\right) \langle \exp\left\{- i \left[\phi({\bf r}_1)-\phi({\bf r}_2)\right]\right\}\rangle.
\end{align*}
Under the assumptions of the typical diffusion plates, we model the phase-phase correlation function 
\begin{align}\label{gaussian C_E}
\langle \exp\left\{- i \left[\phi({\bf r}_1)-\phi({\bf r}_2)\right]\right\}\rangle &= \exp(-\frac{|{\bf r}_1-{\bf r}_2|^2}{\sigma^2}), 
\end{align} 
with a Gaussian decay of width $\sigma$ that is amenable to the following analytic treatments.  The relation
\begin{align}\label{eq:sum_CE_zero}
\langle \exp\left\{- i \left[\phi({\bf r}_1)+\phi({\bf r}_2)\right]\right\}\rangle &= 0, 
\end{align} 
that follows from Eq.~\eqref{eq:zeromean}, in conjunction with the assumption that the correlation function depends only on relative distance, will be useful as well.

\support{
Lets have a look at the computation of this quantity.  As a first step we will re-express the coordinates in terms of relative and center of mass like coordinates, specifically defining
\begin{align*}
\bar {\bf r} &= \frac{{\bf r}_1 + {\bf r}_2}{\sqrt{2}} && {\rm and} & {\boldsymbol \delta} {\bf r} &= \frac{{\bf r}_1 - {\bf r}_2}{\sqrt{2}}.
\end{align*}
I added the $\sqrt{2}$ terms to make the Jacobin of the transformation equal to 1.  In this basis we have the replacements
\begin{align*}
{\bf r}_1 &= \frac{\bar {\bf r} +  {\boldsymbol \delta} {\bf r}}{\sqrt{2}} && {\rm and} & {\bf r}_2 &= \frac{\bar {\bf r} - {\boldsymbol \delta} {\bf r}}{\sqrt{2}}.
\end{align*}
Now preparing for the integrals we need the two terms
\begin{align*}
|{\bf r}_1 - {\bf r}_2|^2 &= 2|{\boldsymbol \delta} {\bf r}|^2 && {\rm and} & |{\bf r}_1|^2 + |{\bf r}_2|^2 &= \frac{1}{2}\left[\left(|\bar {\bf r}|^2 + 2\bar {\bf r}\cdot{\boldsymbol \delta} {\bf r} + |{\boldsymbol \delta} {\bf r}|^2 \right) + \left(|\bar {\bf r}|^2 - 2\bar {\bf r}\cdot{\boldsymbol \delta} {\bf r} + |{\boldsymbol \delta} {\bf r}|^2 \right)\right]. \\
&&&&&= |\bar {\bf r}|^2 + | {\boldsymbol \delta} {\bf r}|^2
\end{align*}
to insert into the argument of the Gaussians.  All together the integral in the denominator then becomes
\begin{align*}
\iint |C_E({\bf r}_1,{\bf r}_2;0)|d^2{\bf r}_1d^2{\bf r}_2 &= \iint d^2\bar {\bf r} d^2{\boldsymbol \delta} {\bf r} \exp(-\frac{|\bar {\bf r}|^2}{w^2}) \exp\left[-\left(\frac{2}{\sigma^2} + \frac{1}{w^2}\right)|{\boldsymbol \delta} {\bf r}|^2\right] = \left[\pi D^2\right]\left[\pi \frac{\sigma^2 w^2}{2w^2 + \sigma^2}\right].
\end{align*}
I added the square brackets at the end to make clear that the $\bar r$ integral doesn't even matter since it is common to the numerator and denominator defining $c_E(0)$.  The integral in the numerator is
\begin{align*}
\iint |C_E({\bf r}_1,{\bf r}_2;0)| |{\bf r}_1 - {\bf r}_2|^2 d^2{\bf r}_1d^2{\bf r}_2 &= \left[\pi w^2\right]\iint d^2{\boldsymbol \delta} {\bf r} \left[2 |{\boldsymbol \delta} {\bf r}|^2\right] \exp\left[-\left(\frac{2}{\sigma^2} + \frac{1}{w^2}\right)|{\boldsymbol \delta} {\bf r}|^2\right] = \left[\pi w^2\right]\left[2\pi\left(\frac{\sigma^2 w^2}{2w^2 + \sigma^2}\right)^2\right].
\end{align*}
Finally giving
\begin{align*}
c_E(0)^2 &= \frac{2 w^2 \sigma^2 }{2w^2 + \sigma^2} = \sigma^2\left(1 + \frac{2 w^2}{\sigma^2}\right)^{-1} \approx \sigma^2\left(1 - \frac{2 w^2}{\sigma^2}\right) \xrightarrow[w\rightarrow\infty]{} \sigma^2
\end{align*}
So we see that for for large beam waists, we recover the intrinsic correlation length.  Great!  When it comes to know the correlation function at all positions, this set of variables will continue to be handy!

How shall we proceed, and what to expect?  Well the next task is to compute the correlation function $C_E({\bf r}_1,{\bf r}_2;z)$ at all $z$, from physical arguments, we expect a result of the same mathematical form as in Eq.~\eqref{gaussian C_E}, with a Gaussian width with a new width $D(z)$, and a correlation length with a new scale $\sigma(z)$.  I.e., we expect a solution of the form (up to a complex phase)
\begin{align*}
\frac{C_E({\bf r_1},{\bf r_2};z)}{E_0^2} &\propto \exp(-\frac{{\bf r_1}^2 + {\bf r_2}^2}{w(z)^2}) \exp(-\frac{|{\bf r_1}-{\bf r_2}|^2}{\sigma(z)^2}).  
\end{align*} 
Now that we know the shape of the answer, lets find it!

There are two important terms in the correlation function, and we need to know both for arbitrary $z$ (don't worry the first is trivial!)
\begin{align*}
\langle E({\bf r}; z)\rangle &=  \frac{-i k_0}{2\pi z}\int d^2{\bf r'} \langle E^+({\bf r'}; 0)\rangle e^{-i k_0|{\bf r}-{\bf r'}|^2/2z} = 0
\end{align*}
the next term is less nice, but not {\it too} horrible
\begin{align*}
\langle E({\bf r}_1; z)E({\bf r}_2; z)^*\rangle &= \frac{k_0^2}{4\pi^2 z^2} \int d^2{\bf r}^\prime_1 \int d^2{\bf r}^\prime_2 \langle E({\bf r}_1^\prime; 0)E({\bf r}_2^\prime; 0)^*\rangle \exp\left[{-\frac{i k_0}{2z}\left(|{\bf r}_1-{\bf r}^\prime_1|^2 - |{\bf r}_2-{\bf r}^\prime_2|^2\right)} \right]
\end{align*}
As a sanity check, $C_E({\bf r_1},{\bf r_2};z)$ reduces to $C_E({\bf r_1},{\bf r_2};0)$ when $z=0$.

checked with the modified definition of $z_R$, $w(z)$, $R(z)$ can be expressed in the same way as usual Gaussian beam.

In case one
\begin{equation}
    \begin{cases}
    &1/R(z) = 1/z - k_0^2w^2/8Az^3\\
    &1/\sigma^2(z)=k_0^2w^2/8z^2 - k_0^4w^4/64Az^4 - k_0^2/16Az^2\\
    &1/w^2(z) = k_0^2/8Az^2\\
    &A = 1/2w^2 + 1/\sigma^2 + k_0^2w^2/8z^2
    \end{cases}
\end{equation}
These quasi-Gaussian coefficients are expressed in terms of the effective Rayleigh length(s?) $z_{\rm R}$, $z_{{\rm R}, \sigma}$.  Depending on the equations, these may be related to the usual Rayleigh range by a $M^2$ parameter.
}

We first consider the case illustrated by Fig.~\ref{fig:speckle_model}(a) where a Gaussian beam goes through a large disordered medium. The field-field correlation function at all positions following the disordered medium  can be exactly computed and takes the form 
\begin{align}
\frac{C_E({\bf r}_1,{\bf r}_2;z)}{E_0^2} =& \left[\frac{w}{w(z)}\right]^2 \exp(-ik_0\frac{{\bf r}_1^2 - {\bf r}_2^2}{2 R(z)})\label{eq:C_E}\\
&\times \exp(-\frac{{\bf r}_1^2 + {\bf r}_2^2}{w(z)^2})\exp(-\frac{|{\bf r}_1-{\bf r}_2|^2}{\sigma(z)^2}) \nonumber 
\end{align}
reminiscent of that of Gaussian beams.

This correlation function is characterized in terms of three $z$-dependent functions: the beam waist $w(z)$, the radius of curvature $R(z)$, and the correlation length $\sigma(z)$.  Each of these is simply related to a reduced Rayleigh range $z_{\rm R}^* = z_{\rm R}/M$, with conventional Rayleigh range $z_{\rm R} = k_0 w^2/2$ and beam quality factor $M^2 = 1+2w^2/\sigma^2$.  The resulting coefficients
\begin{align}
    \left[\frac{w(z)}{w}\right]^2 &= \left[\frac{\sigma(z)}{\sigma}\right]^2 = 1+\left(\frac{z-z_0}{z_{\rm R}^*}\right)^2\label{eq:rayleigh}
\end{align}
and
\begin{align}
    \frac{R(z)}{z-z_0} = 1 +\left(\frac{z_{\rm R}^*}{z-z_0}\right)^2
\end{align}
take the same form as a usual Gaussian beams focused at $z_0$.  Lastly, as in Fig.~\ref{fig:speckle_model}(b), an ideal lens with focal length $f$ at position $z_L$ gives new Gaussian beam parameters defined by
\begin{align}\label{lens making}
\frac{w^\prime}{w} &= \frac{\sigma^\prime}{\sigma} = f \left[\left(z_0'-z_L-f\right)^2+z_{\rm R}^{*2}\right]^{-1/2}
\end{align}
and
\begin{align*}
\left(z_0^\prime-z_L\right)^{-1} &=  f^{-1} - \left[\left(z_L-z_0\right) + \frac{z_{\rm R}^{*2}}{z_L-z_0-f} \right]^{-1}
\end{align*}
\support{
results from the paper, for large $z_{\rm R}^*$, $w'$ is complex. $z_0'$ agrees.
\begin{align}
\frac{w^\prime}{w} = \frac{\sigma^\prime}{\sigma} &= \left[1 + \left(\frac{z_L-z_0}{f} \right)^2 - \left(\frac{z_R^*}{f}\right)^2\right]^{-1/2}, & {\rm and} && \left(z_0^\prime-z_L\right)^{-1}  &=  f^{-1} - \left[\left(z_L-z_0\right) + \frac{(z^*_R)^2}{z_L-z_0-f} \right]^{-1},
\end{align}
}
where the first expression defines the magnification and the second is analogous to the usual lens makers equation~\cite{Self1983}.  While this leaves $M^2$ unchanged, the Rayleigh range is altered owing to the change in $w$.  
All together these relations fully define field-field correlation function $C_E$ throughout an ideal imaging system. 

\support{
Lets have a go at simplifying $\sigma^2(z)$.
\begin{align*}
 \sigma^2(z)&=2w^2\frac{z^2+(z_R^*)^2}{z_R^2-(Z_R^*)^2} = 2w^2\frac{1+z^2/(z_R^*)^2}{z_R^2/(z_R^*)^2-1} = 2w^2\frac{1+z^2/(z_R^*)^2}{M^2-1}= 2w^2\frac{1+z^2/(z_R^*)^2}{2w^2/\sigma^2}\\
 &= \sigma \left(1+z^2/(z_R^*)^2\right).
\end{align*}
Remarkable: Even $\sigma$ takes same form as for the waist.  So I will simplify the equation to take this structure instead.
}

In most quantum-gas experiments, optical potentials are created using laser light in the far detuned limit, thereby experiencing a potential proportional to the optical intensity
\begin{equation}\label{intensity}
    I({\bf r}; z) = \frac{c \epsilon_0}{2} \left|E({\bf r}; z)\right|^2
\end{equation}
not the electric field directly.  The ensemble-averaged intensity 
\begin{align}
\langle I({\bf r}; z) \rangle &= \frac{c \epsilon_0}{2} C_E({\bf r},{\bf r}; z), \label{eq:intensity}
\end{align}
simply related to the field-field correlation function in Eq.~(\ref{eq:C_E}), contains no information about the optical speckle excepting for the changed $M^2$.

As discussed in the next section, the power spectral density (PSD) of the intensity
\begin{align}
    \rho({\bf k}; z) &= \langle \tilde{I}({\bf k};z)\tilde{I}^*({\bf k};z)\rangle\nonumber \\
    &=  \frac{\pi^2w^2(z)}{4M^2}\exp{-\frac{{\bf k}^2w^2(z)}{4M^2}},\label{psd}
\end{align}
computed using Eq.~\eqref{eq:C_E}, describes the momentum-change imparted by the speckle potential to a moving atomic wavepacket.  

\subsection{Correlation length}
The field-field correlation length
\begin{align}
    c_E(z)^2 &= \frac{\iint |C_E({\bf r}_1,{\bf r}_2; z)||{\bf r}_1-{\bf r}_2|^2d^2{\bf r}_1d^2{\bf r}_2}{\iint |C_E({\bf r}_1,{\bf r}_2 z)|d^2{\bf r}_1d^2{\bf r}_2}\\
    &= \frac{2 w(z)^2 \sigma(z)^2}{2w(z)^2 + \sigma(z)^2} \approx \sigma(z)^2
\end{align}
obtained from Eq.~(\ref{eq:C_E}), sets the scale over which the electric field retains its spatial coherence.  The field-field correlation length is minimized at $z=z_0$, and is always larger than $\sigma$.  Generally speckle beams operate in the regime $w \gg \sigma$, where there are many speckle grains within a large beam, giving the final approximate relation. 

As was already noted in our ray-optics discussion, this has important implications for experiment design.  For cold atom experiments such as ours, the large momentum-change imparted by short-length scale speckle is essential, where a correlation length at or below the micron scale is desirable.  Since the correlation length available for typical commercial diffusers ranges form $10\ {\rm \mu m}$ to $100\ {\rm \mu m}$, an additional focusing stage is required .  

A focusing lens can easily take the $10\ {\rm \mu m}$ to $100\ {\rm \mu m}$ correlation length available for typical commercial diffusers and create a beam with sub-micrometer correlation length at its focus.  Figure~\ref{fig:speckle_model}c compares the correlation length of a beam with (red solid) and without (red-dashed) a focusing lens for the specific case of an initial laser beam of wavelength $\lambda = 532\ {\rm nm}$ with input Gaussian beam parameters: focal point $z_0=0$, beam waist $w = 25\ {\rm mm}$ and correlation length $\sigma = 100\ \mu{\rm m}$.  
This beam is focused by a lens of focal length $f=100\ {\rm mm}$, the correlation length at the focus is $c_E=0.96\ \mu{\rm m}$.  The remaining derived beam parameters are $M^2 \approx 1.25\times10^5$, $z_R \approx 3.7\ {\rm km}$, and $z_R^* \approx 10.4\ {\rm m}$. 

\subsection{Impact of apertures}

In the case of focusing optical speckle as shown in Fig.~\ref{fig:speckle_model}(b), a lens of focal length $f$ and diameter $D_L \ll w$ is placed at $z=z_L \leq k_0\sigma^2$. The field in the plane $z=z_L$ before the lens, $E^-({\bf r};z_L)$ is essentially unchanged from field $E^+({\bf r};0)$. The field $E^-({\bf r};z_L)$ passes through the lens aperture, where it acquires a position dependent phase and is truncated outside the lens. The emerging field $E^+({\bf r};z_L)$ propagates to the focal plane $z=f+z_L$ where it is 
\begin{equation}\label{field in z}
    E_f({\bf r}) = \frac{-i k_0}{2\pi f}e^{-i k_0{\bf r}^2/2f}\int \displaylimits_{|{\bf r'}|<\frac{D_L}{2}} d^2{\bf r'} E^+({\bf r'}; 0) e^{i k_0{\bf r}\cdot{\bf r'}/f}.
\end{equation}
When $\sigma \ll D_L \ll w$, the field-field correlation function at the focal plane is
\begin{align}
C_{E,f}({\bf r}_1,{\bf r}_2) \approx & C_0\exp[-\frac{ik_0({\bf r}_1^2-{\bf r}_2^2)}{2f}]\\
&\times\exp[\frac{-k_0^2\sigma^2({\bf r}_1 + {\bf r}_2)^2}{16f^2}]\frac{J_1(k_c \Delta r/2)}{k_c\Delta r /2}.\nonumber
\end{align}
Here $C_0 =  k_0^2E_0^2D_L^2\sigma^2 / 8f^2$ is the peak correlation amplitude; $\Delta r = |{\bf r}_1-{\bf r}_2|$ is the relative position coordinate; and $J_1$ is a Bessel function of the first kind.  The ratio 
\begin{align}
k_c &= k_0\frac{D_L}{f}\label{kc}
\end{align}
is a cutoff above which the PSD of the intensity
\begin{align}
    \rho_f(k) = C_0^2 \frac{2}{\pi k_c^2}\left[ \cos^{-1}\left(\frac{k}{k_c}\right)
     -\frac{k}{k_c}\sqrt{1-\frac{k^2}{k_c^2}}\right]\label{psd_image}
\end{align}
is strictly zero.  Equation~\eqref{psd_image} is valid near the optical axis where $|{\bf r}_1|,|{\bf r}_2| \ll w(z)$.

\subsection{Field and intensity probability distribution}

In the previous sections, we focused on the average properties of speckle fields.
Here we extend this discussion to predict the probability distribution of the electric field strength $P(E)$ and intensity $P(I)$.
Our approach focuses first on $P(E)$, and consists of two steps: (1) we find the regime when the central limit theorem applies, thereby assuring a Gaussian probability distribution; and (2) we identify $\langle E \rangle$ and $\langle E^2 \rangle$ as the lowest moments of the distribution, fully defining the Gaussian distribution.

We now interpret the electric field
\begin{align*}
    E({\bf r}; z) &= \frac{-i k_0}{2\pi z}\int d^2{\bf r'} E^-({\bf r}')e^{-i \phi({\bf r'}) } e^{-k_0|{\bf r}-{\bf r'}|^2/2z},
\end{align*} 
of Eq.~\eqref{para field} as a random variable constructed from a sum over incoherent complex phasors.  The cross correlation function (CCF) $\langle E({\bf r}_1; z) E({\bf r}_2; 0)\rangle$ specifies the range over which the initial random field contributes to the final field.  The closed form expression for this CCF is similar to the field-field correlation function in Eq.~\eqref{eq:C_E}; the length scale for the decay of correlations $\sigma_{\rm CCF}(z)$ again obeys  Eq.~\eqref{eq:rayleigh}, but with $M_{\rm CCF}^2=(1 + w^2/\sigma^2 )^2$.  When $w\gg\sigma$, i.e., the initial waist is much larger than the speckle size,  the resulting Rayleigh range reduces to $z_{\rm R, CCF} = k_0 \sigma^2/2$: as if each random source was an individual Gaussian beam with extent $\sigma$.  The criterion that a field $E({\bf r}; z)$ have contributions from many incoherence sources is therefore $\sigma_{\rm CCF}(z)/\sigma\gg1$, i.e., $z\gg z_{\rm R, CCF}$.  

This identifies the central limit theorem's regime of applicability, and we now consider $E({\bf r}; z)$ as a complex valued Gaussian random variable.  The probability distribution for electric field is therefore a function of two independent degrees of freedom, here we select the quadrature variables $E$ and $E^*$, giving $P(E, E^*)$.   Most moments of this quantity are easy to identify using Eqs.~\eqref{para field}, \eqref{eq:zeromean}, \eqref{gaussian C_E} and \eqref{eq:sum_CE_zero}: $\langle E \rangle = \langle E^2 \rangle = 0$, and similarly for $E^*$.   Then Eqs.~\eqref{C_E} and the following discussion assure us that  $\langle E E^* \rangle = \langle |E|^2 \rangle$ takes on a non-zero value.  Together these fully define the Gaussian probability distribution for electric fields
\begin{align}
    P(E, E^*) &= \frac{1}{\pi \langle |E|^2 \rangle} \exp\left(-\frac{|E|^2}{\langle |E|^2 \rangle}\right)\label{eq:dist_fields},
\end{align}
and using Eq.~\eqref{eq:intensity}, the intensity distribution
\begin{align}
    P(I) &= \frac{1}{\langle I \rangle} \exp\left(-\frac{I}{\langle I \rangle}\right) \label{P of int}
\end{align}
follows directly.
The intensity of a speckle field obeys an exponential distribution and the mean speckle intensity $\langle I \rangle$ should be equal to its standard deviation $\sqrt{\langle I^2 \rangle}$.

\subsection{Simulated speckle and the comparison to experiment} \label{Numerical_speckle}
\begin{figure}[tbp]
    \begin{center}
    \includegraphics{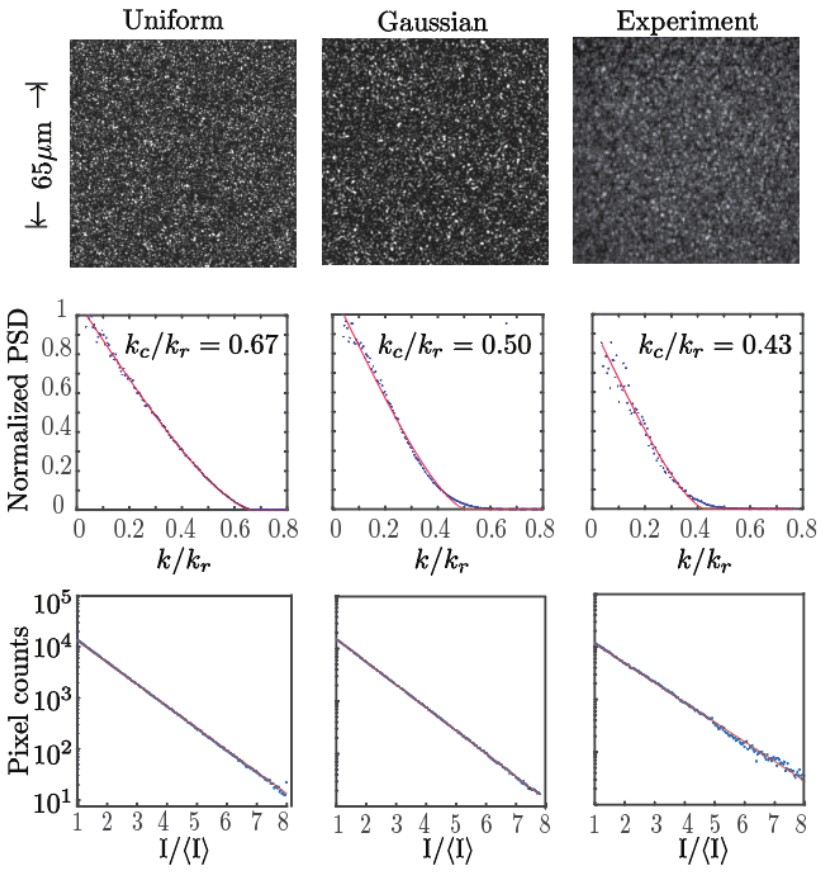}
    \end{center}
    
    \caption{Simulated and measured optical speckle. The columns in the figure correspond to: simulated speckle with uniform laser beam,  simulated speckle from a Gaussian laser beam and measured speckle. In each column, the first row shows the intensity of the optical speckle field. The second row shows the PSD of the intensity shown in the first row (symbols).  The red curve shows a fit of Eq.~(\ref{psd_image}) to the data, along with the resulting $k_c$. The third row histograms the  intensity from the first row.}
    \label{fig:Optical speckle}
\end{figure}

Having now fully set the stage for understanding and creating speckle laser beams, we turn to a laboratory confirmation of key prediction of these models relevant to cold atom experiment: the field-field correlation length $C_E$ and the distribution of intensities $P(I)$.

In our lab, we directed a collimated laser beam (waist $w\approx25\ {\rm mm}$) through a diffuser (divergence angle $\theta_d = 0.5^\circ$, and aperture $D=20\ {\rm mm}$) focused by immediately by a lens (focal length $f = 30\ {\rm mm}$) as depicted by Fig.~\ref{fig:speckle_model} and quantified the the optical speckle formed at the focal plane.
We then imaged the optical speckle onto a charge coupled device (CCD) camera using a Keplerian telescope with magnification $M=46$.
The CCD's $1024 \times 1280$ array of $4.8\ \mu {\rm m}$ pixels gave a $100\ \mu{\rm m} \times 130\ \mu{\rm m}$ magnified field of view with $0.1\ \mu {\rm m}$ pixels.

Our analytic results for $C_E$ are valid in the Gaussian beam limit ( $w\ll D$) or uniform illumination limit ($w\gg D$).  Because our experiment has $w\approx D$, we numerically simulated the optical speckle to compare with our measurements and both models.

For the numerical simulation, the desired optical speckle field $E_{i,j}$ is represented by a $1024 \times 1280$ array at the focal point of the lens.  We use the optical Fourier transform property of lenses to compute this efficiently, whereby the field a focal distance beyond the lens is related to the Fourier transform of the field a focal distance prior to the lens (which we will term the Fourier plane).  An important aspect of this method is that the $0.1\ \mu {\rm m}$ grid spacing in the focal plane transforms to a $1.5\ {\rm mm}$ grid spacing in the Fourier plane.

Our simulation progresses as follows.  (1) We first initialize $E_{i,j}(z=0)$ to the field of either a uniform field or a Gaussian beam.  (2) We then imprint random phases on each point~\footnote{The grid size is much larger than the correlation length of the diffuser, so the imprinted phase at each grid point is uncorrelated with all other points.}.  (3) We set the field outside our physical aperture to zero.  (4) Then we back-propagate the field to the Fourier plane and take the Fourier transform to obtain the field at the focal plane.

Figure~\ref{fig:Optical speckle} compares our measured speckle with numerics and our analytic model; the three columns depict: the case of a uniformly illuminated aperture, Gaussian illumination, and experiment.   The top row shows that intensity at the focal plane is qualitatively similar for all three cases. The middle row, the PSD (computed from the intensity in the top row, and plotted by blue symbols), highlights the differences.  In each case, we fit Eq.~(\ref{psd_image}) the PSD and extracted $k_c$ from the fits (red curves).  Because  Eq.~(\ref{psd_image}) was derived for a uniformly illuminated aperture it provides a good fit to the uniform illumination case, but deviates at large $k$ for Gaussian illumination and experiment.  In contrast, the numerics for Gaussian illumination and the experiment are indistinguishable.  
The bottom row, we histogram the intensity distribution and verify that in all three cases we recover the expected exponential fall-off. 

\section{Scattering of an SOBEC from a speckle potential} \label{Model}

\begin{figure}[htbp]
    \centering
    \includegraphics{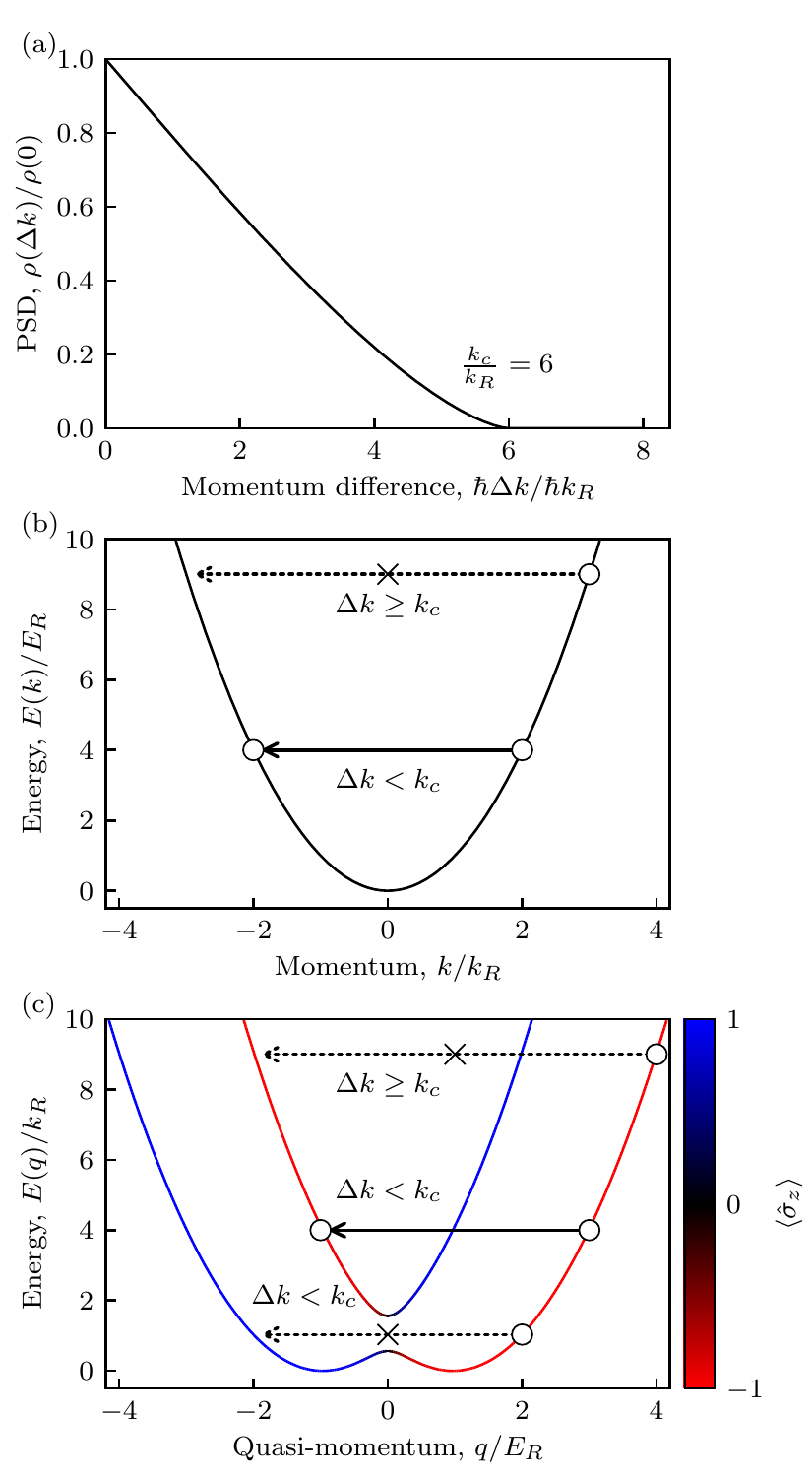}
    \caption{Fermi's Golden Rule. Momentum are expressed in units of the single-photon recoil momentum $\kr$ used to create SOC in (c). (a) Representative PSD for optical speckle with $k_c = 6 \kr$. (b) Free particle dispersion relation. The dashed arrow marks the boundary above which the FGR rate vanishes, while the solid arrow provide an example with non-zero rate. (c) SOC dispersion relations computed for $\delta = 0$ add $\Or = 1\Er$ colored according to the expectation value $\langle\sigma_z(q)\rangle$, with arrows marked as in (b).  Note the transition through the gap in the dispersion relation at $E\approx \Er$ where the FGR rate is nearly zero.}
    \label{fig:Dispersion_relations}
\end{figure}

We now focus on the motion of spin-orbit coupled bosons in a speckle-induced disorder potential.  In this section, we develop a Fermi's golden rule (FGR) approach for scattering from a disorder potential, both with and without SOC, schematically depicted in Fig.~\ref{fig:Dispersion_relations}.   The first order scattering processes captured by the FGR are possible when a matrix element (here from the disorder potential) can couple energetically degenerate initial and final states (here momentum or quasi-momentum states).  We will see that the strength of this coupling is proportional to the PSD of the speckle potential, an example of which is shown in Fig.~\ref{fig:Dispersion_relations}(a).  As depicted in  Fig.~\ref{fig:Dispersion_relations}(b), this implies an absence of scattering for momenta differences larger than the speckle-cutoff $k_c$.  Adding SOC, as in Fig.~\ref{fig:Dispersion_relations}(c), can suppress scattering for additional wavevectors.  Because a spin-independent speckle potential has no spin-changing matrix element, the energetically allowed transition at an energy $E/\Er \approx 1$ between states of opposite spin is strongly suppressed.  The following discussion quantifies these observations.

\subsection{Fermi's golden rule}

In this section we first develop our understanding of scattering from disorder potentials by deriving the FGR for spinless particles.  With that understanding in hand we turn to the case adding SOC.

\subsection{Spinless atoms}

For spinless free particles, the unperturbed Hamiltonian $H = \hbar^2 k^2 / 2m$ implies that we will study scattering between initial and final momentum states, labeled by  $\ket{k_0}$ and $\ket{k_f}$ respectively.  Figure~\ref{fig:Dispersion_relations}(b) depicts examples by open circles, with arrows connecting initial states to final states.

The time evolution of the initial state $\ket{\psi(0)} = \ket{k_0}$ subject to the speckle potential $V(x)$ may always be expressed as
\begin{align} 
\ket{\psi(t)} &= \sum_{k}C_{k,k_0} (t) e^{-i\omega_k t}\ket{k}, \label{expand}
\end{align}
with $C_{k,k_0}(0) = \delta_{k,k_0}$ and $\hbar\omega_k = \hbar^2 k^2 / 2 m$.  The coefficients $C_{k,k_0} (t)$ are governed by the time-dependent Schr\"{o}dinger equation giving the exact expression
\begin{align}
    C_{k_f,k_0}(t) = &  C_{k_f,k_0}(0) + \label{allorder} \\
    & \frac{1}{i\hbar}\sum_k \bra{k_f}\hat V\ket{k} \int_0^t d\tau e^{i\omega_{k_f,k}\tau}C_{k,k_0}(\tau).\nonumber
\end{align}
with 
\begin{align}
    \omega_{k,l} &= \omega_k-\omega_l, &{\rm and} && \hat V &= \sum_x V(x) \ket{x}\bra{x}.
\end{align}
An order-by-order perturbation theory is typically obtained by recursively inserting the integral expression for $C_{k_f,k_0}(t)$ back into the integrand; unfortunately, the general problem is intractable and we truncate the perturbation series at first order. This term is effectivly obtained by replacing $C_{k_f,k_0}(\tau)$ with $C_{k_f,k_0}(0) = \delta_{k_f,k_0}$, and find
\begin{equation}\label{firstC}
    C_{k_f,k_0} (t) = \delta_{k_f,k_0} + \frac{1}{i\hbar}\int_0^t d\tau\bra{k_f}\hat V\ket{k_0} e^{i\omega_{k_f,k_0}\tau}.
\end{equation}
Unfortunately we do not know $V(x)$ for any specific realization of the speckle potential.

In Sect.~\ref{characteristics} we characterized optical speckle in terms of second-order statistical metrics such as the PSD, here equal to $\rho(k_f-k_0) = \langle\bra{k_f}\hat V\ket{k_0}\bra{k_0}\hat V\ket{k_f}\rangle$, where the double-brackets indicate the ensemble average. The resulting ensemble averaged transition probability
\begin{equation}
 P_{f,0} (t) = \frac{\rho(k_f- k_0)}{\hbar^2}\left[\frac{2}{\omega_{f,0}} \sin\left(\frac{\omega_{f,0}t}{2}\right)\right]^2
\end{equation}
is a sharply peaked function centered at $\omega_{f,0} = 0$ with width $2\pi/t$, showing that a narrow range of energy matching states can be populated.  For long times, $\omega_{f,0}t \gg 1$ the quantity in square brackets converges to a scaled Dirac delta function $t\times\delta(\omega_{f,0})$.

Figure~\ref{fig:Dispersion_relations}(a) displays the normalized PSD for a speckle potential computed with $k_c = 6 \kr$, reminding us that $\rho(k) = 0$ for $k\geq k_c$.  Our FGR expression allows two types of scattering processes for the free particle dispersion shown in Fig.~\ref{fig:Dispersion_relations}(b).  In the first process, depicted by the black arrow, the atom's initial momentum is reversed, changed by $\Delta k = 2 k_0$; as indicated by the dashed line, this process is second-order forbidden for $k_0 \geq k_c/2$.  In the second process (not pictured), the atom's momentum is only infinitesimally changed: spreading the wave-packet, but leaving the average momentum unchanged.
This picture shows that back-scattering is essential for momentum-relaxation.

\subsection{Spin-orbit coupled atoms}

Our 1D SOC coupling~\cite{lin2011spin} is created by illuminating a two-level atom with a pair of counter-propagating lasers with wavelength $\lambdar$ tuned to drive stimulated Raman transitions between states $\{ \ket{q+\kr,\uparrow},\ket{q-\kr,\downarrow} \}$.  Here $\hbar\kr = 2\pi\hbar/\lambdar$ and $\Er = \hbar^2 \kr^2 / 2m$ are the single-photon Raman recoil momentum and energy respectively.
Subject to this Raman coupling, the atoms obey the 1D Hamiltonian
\begin{align}
\hat{H}(q) = \frac{\hbar^2 }{2m}\left(q\hat 1 + \kr\hat \sigma_z\right)^2 + \frac{\delta}{2}\hat \sigma_z - \frac{\hbar\Or}{2}\hat \sigma_x,\label{eq:soc}
\end{align}
where $\left\{\hat 1, \hat\sigma_x,\hat\sigma_y,\hat\sigma_z\right\}$ are the identity and Pauli operators, respectively.
Here $q$ is the quasi-momentum, $\Or$ is Raman coupling strength, and $\delta$ is the detuning from the two-photon Raman resonance condition.
The resulting dispersion relations, plotted in Fig.~\ref{fig:Dispersion_relations} for $\delta = 0$ and $\Or = 1\Er$, have energies $E^\pm(q)$  labeled by $q$ along with $\pm$ to indicate if they are in the upper or lower band.

These new energies and their associated amplitudes
\begin{align*}
\ket{q,\pm} &\propto a_\pm(q)\ket{q-\kr,\downarrow} + b_\pm(q)\ket{q+\kr,\uparrow}
\end{align*}
change the potential scattering processes, which we again compute using a FGR expression.  The coefficients
\begin{align*}
a_\pm(q) & = \mp\frac{\Or}{2}& {\rm and} && b_\pm(q) &= \pm\frac{\Delta(q)}{2} + \frac{\sqrt{\Delta^2(q) + \Or^2}}{2}.
\end{align*}
along with the quasi momentum dependent detuning
\begin{align}\label{delta}
    \Delta(q) &= \frac{2\hbar^2q\kr}{m} + \delta
\end{align}
fully define these superposition states.

Following the same FGR argument presented above for initial states $\ket{q_0,-}$ in the lower dispersion scattering from a spin-independent speckle potential, we find scattering probabilities
\begin{align}\label{transfer prob}
    P^\pm_{f, 0}(t) &= \frac{\rho(\Delta q)}{\hbar^2}\left|\frac{2\sin(\omega^\pm_{f, 0}t)}{\omega^\pm_{f, 0}}\!\matrixel{q_f,\pm}{e^{i\Delta q x}}{q_0,-}\right|^2
\end{align}
expressed in terms of the quasimomentum and energy differences $\hbar\Delta q = \hbar q_f-\hbar q_i$ and $\hbar \omega_{f,0}^\pm = E^\pm(q_f) - E^-(q_0)$.
For most initial states $\ket{q_0,-}$, such as the two higher energy states marked in Fig.~\ref{fig:Dispersion_relations}(c), the scattering is essentially unchanged from our spinless example, with scattering occurring between energy-matched states with the same initial and final spin.
In contrast, for initial states residing in the SOC energy gap there is no energy-matched state of the same spin available for back-scattering; as indicated by the dashed line, scattering is greatly suppressed.
We note that backscattering is not completely blocked, because the energy matching states $\ket{\pm q_0,-}$ are not spin-eigenstates and do have some spin-overlap.

\subsection{Computed scattering rates}

\begin{figure}[tbp]
    \centering
    \includegraphics{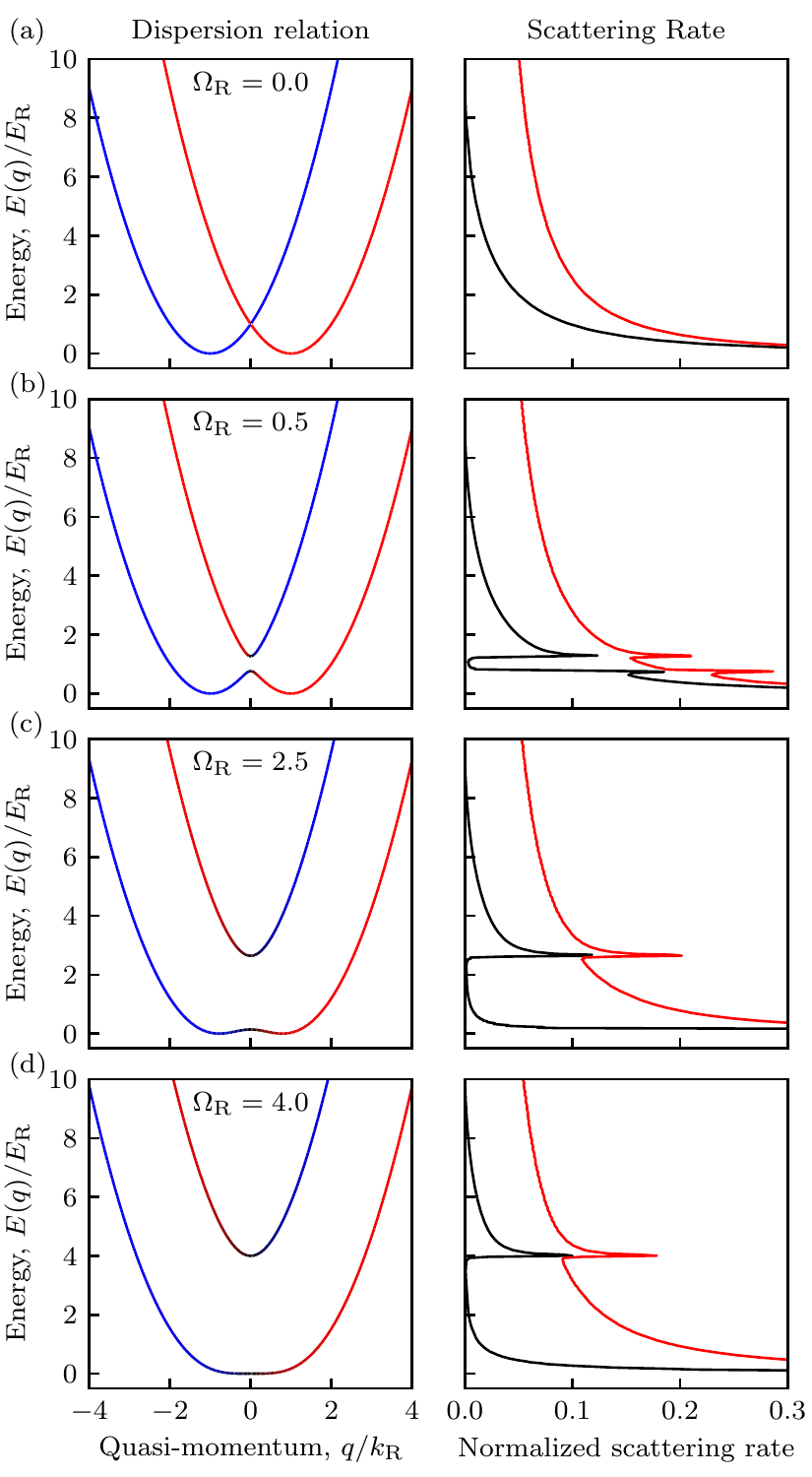}
    \caption{Fermi's golden rule scattering rate for $\Or/\Er = 0, 0.5, 2.5$ and $4.0$.
    Left column: SOC dispersion relations computed for each $\Or$, colored as in Fig.~\ref{fig:Dispersion_relations}.
    Right column: normalized scattering rate as a function of initial energy for the initial state $\ket{q_0, -}$ with $q_0 \geq q_{\rm min}$, i.e., in the bottom dispersion and to the right of the higher momentum local energy minimum.
    }
    \label{fig:scatteringRate}
\end{figure}

We now use these FGR expressions to compute the scattering rates for both forward scattering and back scattering processes.  Because we are interested in transport properties, we define forward scattering processes as those that leave the sign of the group velocity unchanged and back scattering processes and those that do reverse the direction of motion.
We therefore consider initial states $\ket{q_0,-}$ in the lower band with positive group velocity.  Because the lower energy SOC dispersion plotted in Fig.~\ref{fig:scatteringRate} can have a pair of minima located at $\pm q_{\rm min}$, we always select $q_0 > q_{\rm min}$ to assure positive group velocity.  We numerically evaluated the FGR for $^{87}{\rm Rb}$ atoms illuminated with $\lambdar=790\ {\rm nm}$ Raman lasers, giving $\Er = h \times 3.7\ {\rm kHz}$, and for speckle with $k_c = 6 \kr$.  The $t=13.4\ {\rm ms}$ interaction time was selected to be experimentally relevant.  

The right panels of Fig.~\ref{fig:scatteringRate} show the normalized scattering rate computed for four different values of $\Or$, with the back-scattering rate plotted in black and forward scattering plotted in gray.  These rates combine the contributions from the $\pm$ bands in Eq.~(\ref{transfer prob}).

Panel (a), computed for $\Or=0$ (equivalent to the case with no SOC), shows two key effects.  Firstly, the diverging forward and back scattering rates at low energy follow from the diverging density of states (DoS) in 1D.  Secondly, the rate of back scattering (red) falls to zero when $\delta q > k_c$, while forward scattering (black) simply falls with the DoS.  Panels (b) and (c) show cases with a well resolved SOC energy gap.  As expected, back scattering is nearly completely suppressed for initial energies in the energy gap, while forward scattering is hardly changed.  In addition, a pair of singular features boarder the energy gap, resulting from the diverging DoS the local extrema of the dispersions.  Panel (d) shows the same phenomena, but just as the two minima at $\pm q_{\rm min}$ have merged into a single minimum at $q_{\rm min}=0$.  

We therefore conclude, for non-interacting particles back scattering and momentum relaxation is nearly completely suppressed for atoms starting in the SOC energy gap.


\section{Numerical simulation of GPE}\label{GPE_simulation}

Our single particle FGR results only describe the short-time scattering from a disorder potential, they cannot describe the full approach to equilibrium.
To bridge the gap between the FGR and the physical system, we need to account for both higher order scattering processes and interparticle interactions.
In our proposed SOBEC realization, all aspects of SOC Hamiltonian and the speckle potential are tunable, making SOBECs an ideal system for exploring enhanced transport in 1D quantum wires.
\newline
\subsection{Gross-Pitaevskii equations}
Here we numerically study the deceleration of an SOBEC initially moving in a speckle potential using the time-dependent Gross-Pitaevskii equation (GPE).
The time-dependent GPE 
\begin{align}\label{gpe}
i\hbar\partial_t\Psi({\bf r},t) &= \left[-\frac{\hbar^2}{2m}\nabla^2 + V({\bf r}) + g_{\rm 3D}|\Psi({\bf r},t)|^2\right]\Psi({\bf r},t)
\end{align}
is a non-perturbative dynamical description ~\cite{erdHos2010derivation} of a large number of interacting identical bosons occupying the same spatial mode $\Psi({\bf r}, t)$, normalized to the total atom number, $N=\int d^3{\bf r} |\Psi({\bf r}, t)|^2$.
The interaction strength $g_{\rm 3D} = 4\pi\hbar^2a_s/m$ can be expressed in terms of the $s$-wave scattering length $a_s$.
This GPE provides a good description of low-temperature spin-polarized BECs, with negligible thermal excitations~\cite{Dalfovo1999}. 

Since our focus is on 1D transport, we must first obtain a 1D description of our 3D system~\cite{bao2003numerical}.
Here we, we assume that the potential $V({\bf r}) = V(x) + V_\perp(y,z)$ can be separated into a weak longitudinal potential $V_\parallel(x)$ along with a strongly confining transverse potential $V_\perp(y,z)$.
When the single-particle energy spacing from $V_\perp(y,z)$ greatly exceeds all other energy scales, the 3D wavefunction can be factorized into
\begin{equation}
    \Psi({\bf r},t) = \psi(x,t)\phi(y, z),
\end{equation}
containing a longitudinal term of interest giving the 1D density $n(x) = |\psi(x,t)|^2$, and a transverse term $\phi(y, z)$, normalized to unity, assumed to be the ground state of the transverse potential. 
Inserting this ansatz into the GPE and integrating out the transverse degrees of freedom, gives the 1D GPE
\begin{equation}
    i\hbar\partial_t\psi = \left[-\frac{\hbar^2}{2m}\partial_x^2 + V(x) + g|\psi|^2\right]\psi,
\end{equation}
suitable for studying single-component 1D bosons in a speckle potential with 1D interaction strength 
\begin{equation}
    g = g_{\rm 3D}\int dy dz |\phi(y,z)|^4.
\end{equation}
For compactness of notation, here and below, we shall omit the functional dependance of $\psi$ on $x$ and $t$.

\begin{widetext}
The two component 1D spinor GPE describing SOBECs extends Eq.~\eqref{eq:soc} to include interactions, and consists of a pair of coupled non-linear differential equations
\begin{align}
i\hbar\partial_t\psi_\uparrow &= \left[\frac{\hbar^2}{2m}\left(-i\partial_x + \kr\right)^2 + \frac{\delta}{2} + V(x) + g_{\uparrow\uparrow} |\psi_\uparrow|^2 + g_{\uparrow\downarrow}|\psi_\downarrow|^2\right]\psi_\uparrow + \frac{\Or}{2}\psi_\downarrow \\
i\hbar\partial_t\psi_\downarrow &= \left[\frac{\hbar^2}{2m}\left(-i\partial_x - \kr\right)^2 - \frac{\delta}{2} + V(x) + g_{\downarrow\downarrow} |\psi_\downarrow|^2 + g_{\uparrow\downarrow}|\psi_\uparrow|^2\right]\psi_\downarrow + \frac{\Or}{2}\psi_\uparrow
\end{align}
including the interaction strengths $g_{\uparrow\uparrow}$, $g_{\uparrow\downarrow}$, and $g_{\downarrow\downarrow}$.
Here we focus on the specific case of $^{87}{\rm Rb}$ atoms~\cite{kawaguchi2012spinor} in the $f=1$ ground state manifold and have selected $\ket{\uparrow} = \ket{m_F=0}$ and $\ket{\downarrow} = \ket{m_F = -1}$. 
The interactions can be parameterized in terms of an $s$-wave pseudo-potential $(g_{0, {\rm 3D}} + g_{2, {\rm 3D}}\Vec{F}_\alpha\cdot \Vec{F}_\beta) \delta({\bf r}_i - {\bf r}_j)$ now dependent on spin.
In $^{87}{\rm Rb}$'s $f = 1$ manifold $g_{0, {\rm 3D}} = 100.86 \times 4\pi\hbar^2 a_{\rm B}/m$ is vastly larger than $g_{2, {\rm 3D}}\approx-4.7\times10^{-3} \times g_{0, {\rm 3D}}$, where $a_{\rm B}$ is the Bohr radius~\cite{vanKempen2002,Widera2006}.
The interaction coefficients reduce to effective 1D interaction strengths just as in the single component case, and are related to the generic coefficients~\cite{Ho1998,Ohmi1998} via $g_{\uparrow\uparrow} = g_0$ and $g_{\downarrow\downarrow}  = g_{\uparrow\downarrow} = g_0 + g_2$.
Table \ref{table:1} summarizes the parameters used in our simulations. 
\end{widetext}

Our simulation results are divided into two sections: Sec.~\ref{single} hones our understanding by considering a single-component BEC evolving in a speckle potential, and then in Sec.~\ref{soc} we contrast to the case with SOC. 
In both sections, we simulate initially trapped BECs accelerated to an initial momentum $k_0$ or quasi-momentum $q_0$ and we study their deceleration. 
All the results are averaged over 20 speckle realizations, as in  Fig.~\ref{fig:Optical speckle}(b). 
The average speckle potential $h\times 200\ {\rm Hz} \approx 0.05\Er$ was selected to be weak enough to cause no trapping effect yet strong enough to produce significant deceleration within $15\ {\rm ms}$.

\begin{table}
\renewcommand{\arraystretch}{1.25}
\begin{tabular}{l c c}
\hline\hline
Description & Symbol & Value \\
\hline
$^{87}{\rm Rb}$ atomic mass & $m$ & $1.42\times10^{-25}~{\rm kg}$\\
Raman laser wavelength & $\lambdar$ & $790~{\rm nm}$ \\
Recoil energy & $\Er$ & $h\times 3.678\ {\rm kHz}$\\
Dipole trap frequency & $\omega/2\pi$ & $10~{\rm Hz}$  \\
Angle of Raman beams & $\theta_{\rm R}$& $180^{\circ}$ \\
Speckle potential cut off &  $k_c$ & $6\kr$\\
Average speckle potential & $\overline{V(x)}$ & $0.05\Er$\\
Grid spacing  & $\delta x$ & $66\ {\rm nm}$\\
Grids points (single-component)& $N_x$ & $2^{14} + 1$\\
Grids points (SOC)& $N_x$ & $2^{13} + 1$\\
Atom number & $N$ & $2\times 10^5$ \\
Chemical potential & $\mu$ & $h\times 300\ {\rm Hz}$ \\
\hline\hline
\end{tabular}
\normalsize
\caption{Simulation parameters}
\label{table:1}
\end{table}

\subsection{Single component systems}\label{single}

\begin{figure}[tbp]
    \centering
    \includegraphics{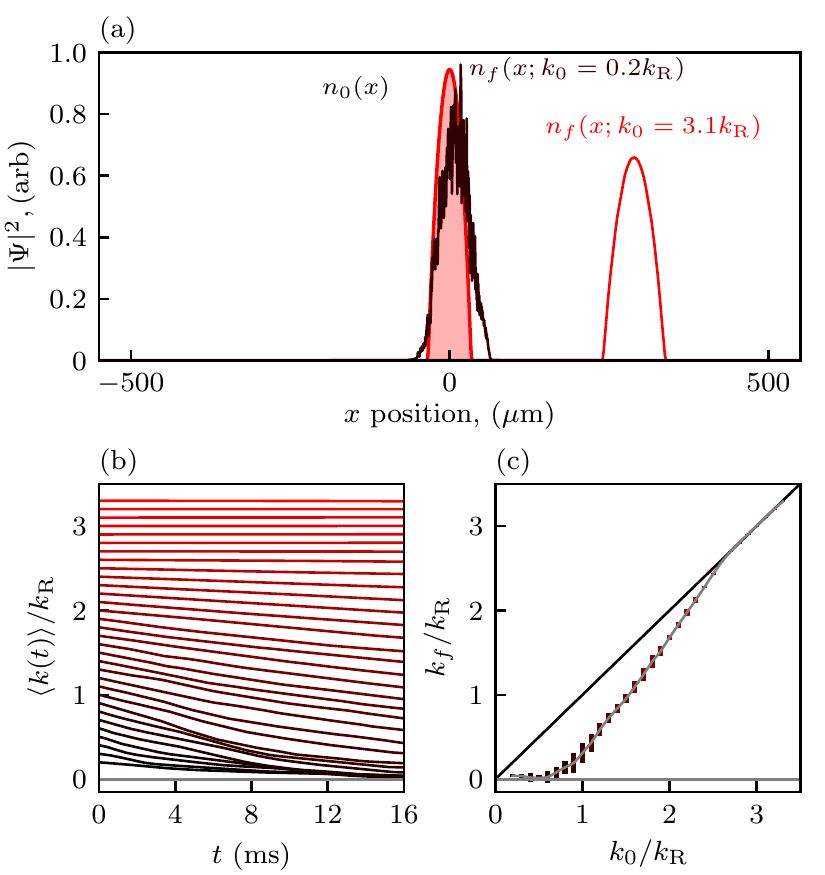}
    \caption{Single-component GPE simulation with $k_c / \kr = 6$.  
    (a) Density distributions.  The filled red curve depicts the initial density distribution, while the black and red curves show the final-state density distributions for initial momenta $\ket{k_0=0.2\kr}$ and $\ket{k_0=3.1\kr}$,above and below $k_c/2$, respectively.
    (b) Mean momentum. $\langle k(t) \rangle$ averaged over 20 speckle realizations is plotted for a range of initial momentum in the range of $0$ to $3.3\kr$, the $t=0$ point of each curve marks the initial $k_0$. 
    (c) Deceleration.  The colored symbols plot $k_f = \langle k(t=16\ {\rm ms}) \rangle$ as a function of $k_0$ along with their standard deviations, and the black line marks $k_f = k_0$ corresponding to ballistic motion.
    }
    \label{fig:single}
\end{figure}

The simulations are performed in three steps to as accurately as possible model a realistic experimental sequence. 
First, we initialize a ground state BEC in a harmonic trap using imaginary time evolution~\cite{chiofalo2000ground}, giving the density distribution plotted in black in Fig.~\ref{fig:single}(a)], and follow with real-time evolution.
Second, because the BEC's narrow momentum distribution is centered at $k=0$, we briefly apply a linear potential $\alpha x$ with time-evolution approximately described by the phase factor $\exp(i k_0 x)$, a momentum translation operator that transforms $\ket{k=0}$ to $\ket{k_0}$. 
Third, having prepared our $\ket{k_0}$initial state, we replace the harmonic potential with a speckle potential (with $k_c/\kr = 6$) and follow the time evolution for $16\ {\rm ms}$.
Figure~\ref{fig:single}(b) captures the main result of this section: when $k_0 > k_c / 2$ the time-evolution is almost unchanged by the speckle potential, while slowly moving initial states are both decelerated and exhibit considerable interference.

Figure~\ref{fig:single}(b) plots the ensemble-averaged momentum $\langle k(t)\rangle$ as a function of time for a range of initial states with $k_0$ from near-zero to $k_0 / \kr = 3.3$, and Fig.~\ref{fig:single}(c) plots the final momentum $k_f$ as a function of $k_0$.
At $t=0$, the average momentum is $\langle k(t)\rangle = k_0$; for $k_0\gtrsim k_c/2$ the BEC evolves ballistically, leaving $\langle k(t)\rangle$ unchanged, while $\langle k(t)\rangle = k_0$ falls rapidly for smaller $k_0$.
Both of these observations are consistent with our FGR analysis which showed a complete absence of momentum changing back-scattering for $k_0 \geq k_c/2$, and with rapidly increasing back-scattering as $k$ falls to zero.

\subsection{SOBECs}\label{soc}

\begin{figure}[t]
    \centering
    \includegraphics{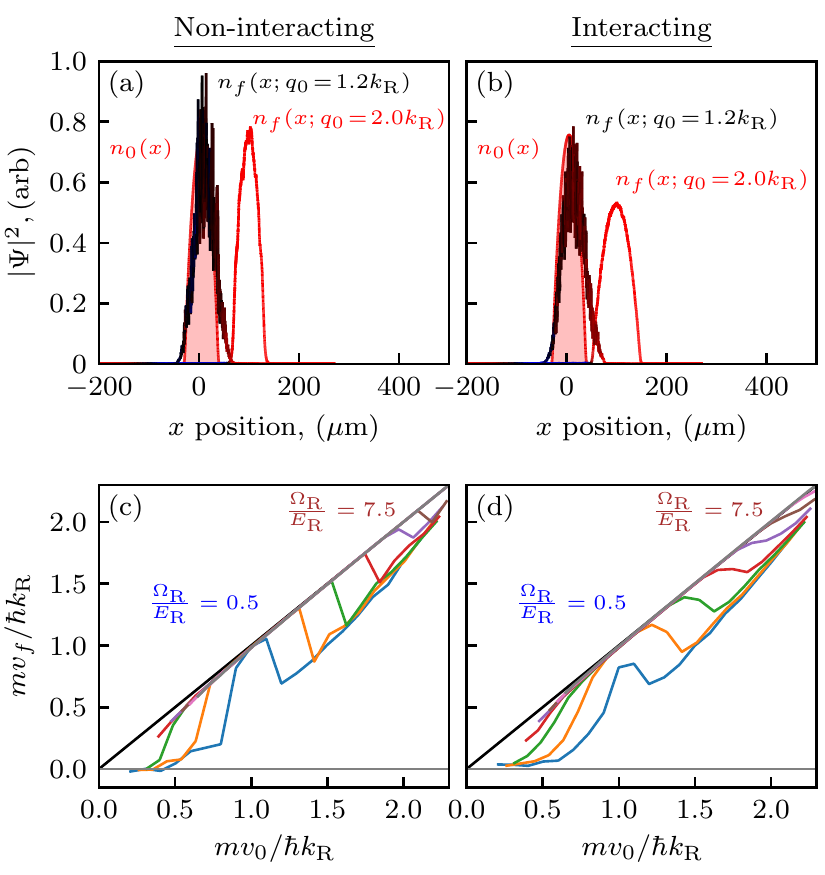}
    \caption{Motion in the presence of speckle and SOC.
    The left column was computed without interactions and the right column added interactions. 
    (a) and (b) Density distributions colored by their magnetization according to the color bar in Fig.~\ref{fig:Dispersion_relations}.
    The shaded curve depicts the initial density distribution, while the remaining red and and black curves were computed for  $q_0=2.0k_{\rm R}$ (in the SOC gap) and $q_0=1.2k_{\rm R}$ (below the SOC gap), respectively.
    (c) and (d) Ensemble averaged final group velocity plotted as a function of initial group velocity for coupling strengths  from $0.5E_{\rm R}$ to $7.5E_{\rm R}$, spaced by $1.0E_{\rm R}$.
    The results in (c) and (d) were averaged over 20 random speckle realizations. 
       }
    \label{fig:SOC}
\end{figure}

As in the single-component case, simulations with SOC begin with three steps aligned with experiment, however, the process of preparing the initial quasimomentum state $\ket{q_0,-}$ is considerably more elaborate than preparing a momentum state $\ket{k_0}$ in a single component system.
(1) As before, we initialize a ground state BEC in a harmonic trap using imaginary time evolution, spin polarized in state $\ket{k_0 = 0, \downarrow}$.
(2) We then use a combination of adiabatic and unitary evolution (described below) to transform this state into $\ket{q_0,-}$ for $\delta=0$ and $\Or$ ranging from $0.5\Er$ to $8\Er$.
(3) Lastly, we again remove the harmonic potential and again follow the time evolution with a speckle potential ($k_c/\kr = 6$) for $16\ {\rm ms}$.

Our procedure (2) begins with the observation that in a frame moving with velocity $\hbar \delta k / m$, the detuning $\delta$ present SOC Hamiltonian in Eq.~\eqref{eq:soc} is Dopper-shifted~\cite{Cheuk2012, Valdes-Curiel2017} to $\delta + 2\hbar^2 \delta k \kr/m$.
Our first task is to adiabatically transform the initial state $\ket{k_0 = 0\downarrow}$ into $\ket{q_{\rm min}, -}$, a ground state SOBEC with quasi-momentum centered at $q=q_{\rm min}$, the global minima of the SOC dispersion, but with $\delta = 2\hbar^2 (q_0 - q_{\rm min}) \kr/m$.
We achieve this by ramping up the Raman coupling strength from zero to $\Or$ on a time scale slow compared to $\hbar/\Delta(q_0,0)$.
In the slow ramp up process, the harmonic trap provides the restoring force required to keep the state at a local minima of the dispersion~\cite{lin2011spin}, i.e., with zero group velocity. 
Lastly, we diabatically set $\delta=0$ and apply momentum kick $\exp[i(q_0-q_m)x]$, giving the desired state $\ket{q_0, -}$. 

Figures~\ref{fig:SOC}(a) and (b) show representative density distributions $n(x) = |\psi_\uparrow(x,t)|^2 +|\psi_\downarrow(x,t)|^2$ before and after a $16\ {\rm ms}$ time evolution with $\Or = 2\Er$, both (a) with no interactions and (b) with interactions.
In both cases the pink shaded curve depicts the initial density distribution, while the density distributions for initial quasimomemta of $q_0 = 1.2 \kr$ and $2.0\kr$ are shown by the black and red curves respectively.
In both cases the momentum exchange for back-scattering is below $k_c$, however, as with the FGR results, these direct simulations show that initial states prepared with energy within the SOC energy gap experience negligible change in velocity, independent of the presence of interactions.

While the free particle group velocity is simply related to wave-vector by $v = \hbar k / m$, atoms evolving according to the SOC dispersions, as in Fig.~\ref{fig:scatteringRate}, have group velocity given by the more complex relation
\begin{align}
\frac{v_\pm}{\vr} &= \frac{q}{\kr} \left\{1 \pm \left[\left(\frac{q}{\kr}\right)^2 + \left(\frac{\Or}{4 \Er},\right)^2\right]^{-1/2}\right\},
\end{align}
for atoms in state $\ket{q, \pm}$, expressed in units of the recoil velocity $\vr = \hbar \kr/m$.
Because we are interested in transport phenomena, it is this group velocity not the quasimomentum, that is the quantity of primary interest.

Figures~\ref{fig:SOC}(c) and (d) plot the final group velocity $v_f$ as a function of initial group velocity $v_0$ after a $16\ {\rm ms}$ period of free evolution, both (c) with no interactions and (d) with interactions, and with $\Or$ from $0.5\Er$ to $7.5\Er$.
As compared to the simulations without SOC in Fig.~\ref{fig:SOC}(c), these curves show a near-complete suppression of relaxation for velocities near $v_0 \approx \vr$, in the SOC energy gap, and with an increasing window of suppression with increasing $\Or$.

Lastly, we see that interaction effects do play a role, leading to more rapid deceleration.
The origin of this effect can be understood by comparing the red curves in Figs.~\ref{fig:SOC}(a) and (b): adding interactions leads a mean-field driven expansion of the BEC, increasing the range of velocities present.
As a result, when the SOC energy gap is small (small $\Or$), a significant fraction of the BEC's velocity distribution falls outside the SOC energy gap, thereby sampling points in the dispersion where first-order backscattering is allowed.
At larger $\Or$, motion is near-ballistic near the center of the SOC gap, but the transition from ballistic to decelerated is smoothed as compared to the case with no interactions.

\section{Conclusion} \label{Conclusion}

Our analytical and numerical studies of the transport of SOBECs in disorder potentials clearly show dramatically enhanced transport for initial states in the SOC energy gap.  
The enhanced transport described here results from the same physics giving rise to a spin transistor in Ref.~\cite{Mossman2019}, which also relied on a combination if kinematic and matrix-element effects to yield non-reciprocal appearing transport behavior.
In the appendix, we describe an explicit experimental proposal using laser speckle derived from $532\ {\rm nm}$ green laser and an off-the-shelf optical diffuser.
In this proposal, SOC is generated from  a pair of $790\ {\rm nm}$ laser beams intersecting at the atoms, and initial states would be prepared as described above.
The protection from back scattering is independent of quantum statistics: non-interacting fermions would experience a conductivity increased by the factor predicted by the FGR when the Fermi energy resides in the SOC gap.
As with the interacting SOBEC we analyzed, we expect that fermionic systems with moderate interactions would show gains in conductivity, however, the details of this latter case would necessitate future study.

Reference~\cite{Hugel2014} showed that in lattices, the type of 1D SOC in Eq.~\eqref{eq:soc} has the same dispersion as the edge modes of 2D ${\rm Z}_2$ topological insulators~\cite{Kane2005}.
Together with our finding, this indicates that 1D nanowires with SOC either of the Rashba~\cite{Bychkov1984} or linear-Dresselhaus~\cite{Dresselhaus1955} type should provide the same protection to backscattering from spin-independent disorder as would be observed at the edge of a topological insulators.

\section{acknowledgments} \label{acknowledgements}

This work was partially supported by the AFOSRs Quantum Matter MURI, NIST, and the NSF through the PFC at the JQI. 
We benefited from laboratory assistance from E.~M.~Altuntas, C.~J.~Billington, and F.~Salces-Carcoba.
We are grateful for the meticulous reading of our manuscript by the afore noted individuals along with W.~McGehee and G.~H.~Reid.
\newline
\appendix*

\section{Speckle beam design for SOC experiments}\label{speckle_design}

\begin{figure*}
    \centering
    \includegraphics{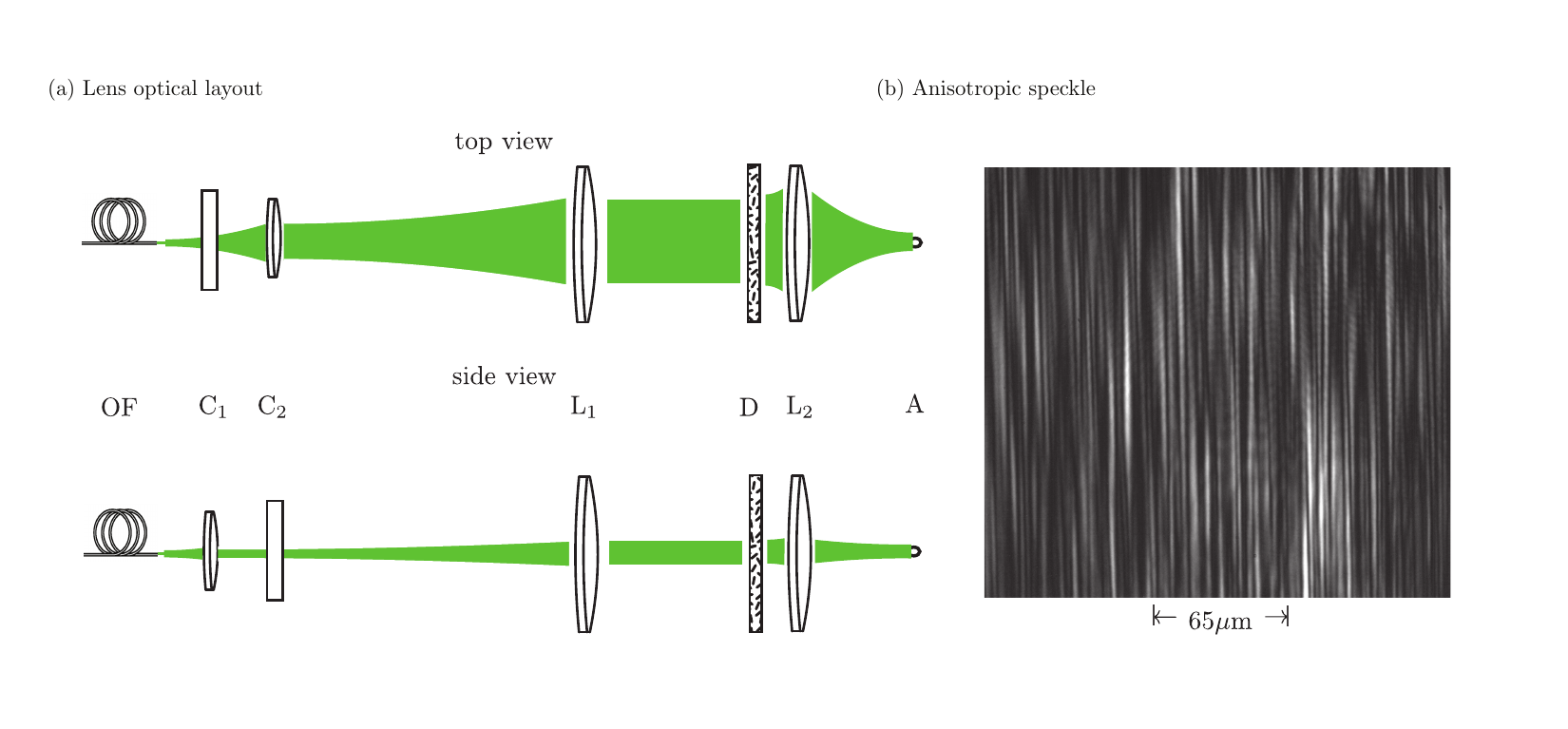}
    \caption{Optical design. (a). The design of optics viewed in two directions. OF denotes optical fiber. Lenses $C_1$ and $C_2$ are cylindrical lenses: $C_1$ focuses the beam in the vertical direction; and $C_2$ focuses the beam in the horizontal direction. $L_1$ is a spherical lens that collimates the beam. D is the optical diffuser that imprints random phase on the beam. $L_2$ is an aspherical lens that focuses the beam to the atoms labelled with A. (b) Experimental image of optical speckle with anisotropic correlation length.}
    \label{fig:design}
\end{figure*}

In practice, the speckle beam must satisfy two requirements.
The first is anisotropic field-field correlation length: small along $\ex$ and large along $\ey$ and $\ez$ so that scattering occurs predominantly along $\ex$
The second is that the beam width along $\ex$ should uniformly illuminate the elongated atomic ensemble (with expected diameter of about $50\ \mu{\rm m}$).
To observe the effect of SOC-suppressed transport, the speckle potential must couple energy matched states across the SOC gap, shown by the dashed line in Fig.~\ref{fig:Dispersion_relations}(c). 
This implies PSD of speckle potential along $\ex$ satisfies $k_c\gtrsim 4\kr$, informing the selection of beam-size and lenses.
The requirement that the correlation length along $\ey$ be large implies that at the diffuser plate, the beam be much smaller along $\ey$ than along $\ex$.

To satisfy these joint requirements, we created the speckle beam shown in Fig.~\ref{fig:design}(a), that begins with a $532\ {\rm nm}$ laser beam emanating from an optical fiber. 
The beam out of an optical fiber travels through the cylindrical lens $C_1$ (focusing along $\ey$) before encountering a cylindrical lens $C_2$ (focusing along $\ex$) as shown in Fig.~\ref{fig:design}(a), given more rapid divergence along $\ex$ than $\ey$.
The beam is then collimated by $L_1$, a $f = 250\ {\rm mm}$ spherical lens, giving a beam width of around $25\ {\rm mm}$ along $\ex$ and less than $500\ \mu{\rm  m}$ along $\ey$ (on the same scale as the diffuser plate's correlation length). 

The beam then traverses the diffuser plate (Edmund Optics~\footnote{Certain commercial equipment, instruments, or materials are identified in this paper in order to specify the experimental procedure adequately.  Such identification is not intended to imply recommendation or endorsement by the National Institute of Standards and Technology, nor is it intended to imply that the materials or equipment identified are necessarily the best available for the purpose.} part number \#47-680, with divergence angle $\theta_d$ is $0.5^\circ$)  and is focused by $L_2$, a $f = 30\ {\rm mm}$ lens.
Figure~\ref{fig:design}(b) shows a test image of speckle beam at the focal plane, its intensity correlation length is less than $0.5\ {\rm \mu m}$ along $\ex$ and about $10\ \mu{\rm  m}$ along $\ey$. 
The beam widths along both directions are about $250\ \mu{\rm m}$.

\bibliographystyle{apsrev4-1} 
\bibliography{bibfile}

\end{document}